# On Quantum Entanglement and Nonlocality

Mafiz Uddin

*Alberta Computational Biochemistry Lab*
*Suite 208, 8909-100 Street, Edmonton, AB, Canada, E-mail: mafiz.uddin36@gmail.com*

# Abstract

EPR showed that two particles emitted from a source can be entangled by a shared wavefunction where two non-commuting observables (position, momentum) can be simultaneously real, leading to a contradiction with quantum mechanics (two non-commuting variables can not be simultaneously real). John Bell derived an inequality where any local hidden variables prediction is bounded and quantum mechanics can violate the inequality. Bell tests on correlated photon pairs showed a clear violation of the Bell inequality and agreement with quantum mechanics. This study revealed that the Bell inequality holds at any individual instances in a given system but it does not hold over the entire population. The Bell inequality is not an incompatibility criterion for local hidden variables vs quantum mechanics rather it is a criterion for an individual nature vs population dynamics. The nonlocality prediction given by wavefunction is a system approximation and local causality (photon polarization) provides a complete description of the Bell tests. The paper concludes with great satisfaction that the measurement of two correlated photons at a distance was a local cause (photon-filter interaction did not violate the special theory of relativity).

# Keywords

EPR Paradox, Bell Theorem, Quantum Mechanics, Entanglement, Nonlocality, Photon Polarization

# Graphical Abstract

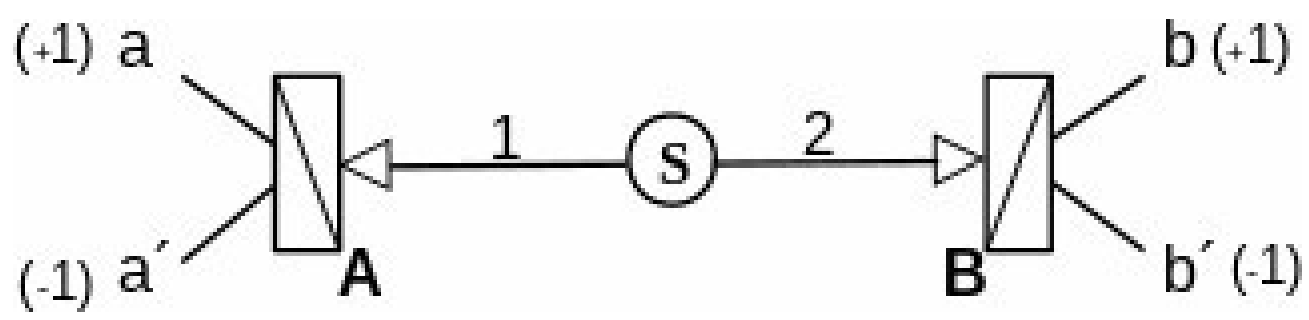

Experimental setup shows that two photons emitted from the same source are interacting with two distance filters. EPR showed that two particles here are entangled by a shared wavefunction where two non-commuting observables (position, momentum) can be simultaneously real, leading to a contradiction with quantum mechanics (two non-commuting variables can not be simultaneously real). EPR argued that the wavefunction description is incomplete because it violates the special theory of relativity (no information can travel faster than the speed of light). This study shows that nonlocal prediction given by wavefunction is a system approximation and local causality photon polarization provides a full description of the experiment.

## 1. Introduction

EPR (Einstein, Podolsky, Rosen) [1] showed that two particles emitted from a source can be entangled by a shared wavefunction where two non-commuting observables (position, momentum) can be simultaneously real, leading to a contradiction with quantum mechanics (two non-commuting variables can not be simultaneously real). EPR argued that the quantum mechanical description of a physical reality given by wavefunction is incomplete because it violates the special theory of relativity (no information can travel faster than the speed of light). There has been a significant interest in possible hidden variable theories and experimental attempts to resolve the EPR argument (*known as the EPR paradox*).

Einstein's acceptable criterion of a reality was: *"If, without in any way disturbing a system, we can predict with certainty (*i.e.*, with probability equal to unity) the value of a physical quantity, then there exists an element of physical reality corresponding to this physical quantity."* Clauser *et al.* [2] quoted Bohr's reply that "*no paradox can be derived from the assumption of completeness if one recognizes that quantum mechanics concerns only the interaction of microsystems with the experimental apparatus and not their intrinsic character.*"

Some arguments have been made to restore new variables to the theory of local causality [3-5]. John Bell derived a theorem stating that any such theory with additional variables will not have complete agreement with quantum mechanics [6]. Clauser and Freedman first conducted Bell tests and drew an affirmative conclusion in favor of quantum mechanics. Aspect *et al.* [7] conducted similar experiments with more refined settings on correlated photon pairs using two-channel polarizers (analogs of Stern-Gerlach filters on pairs of spin 1/2 particles in a singlet state). Their results showed a clear violation of Bell inequality of local hidden variables ($-2 \le S \le 2$) and a complete agreement with quantum mechanics. A team led by Anton Zeilinger conducted the Cosmic Bell test and arrived at the same conclusion in favor of quantum mechanics. The violation of the Bell inequalities in classical optics are now routinely observed.

In the above Bell test, there are four measurement quantities (or four elements of realization). These measured quantities are used to calculate the Bell inequality. In a unified quantum mechanical system, these quantities have no separate realizations. This study will show that, in classical mechanics, we do not have any difficulties to predict these quantities by using Bell's classical definition of local hidden variables (Bell's Equation 2). The study will show that the Bell inequality ($-2 \leq S \leq 2$) holds for any individual instances (i.e. each correlated photon pair) but it does not hold over the entire population in a given system. Nonlocality description given by wavefunction is a system approximation and local causality photon polarization provides a complete description of the experiment. The simultaneous measurement (photon-filter interaction) of the correlated photons at a distance did not violate the special theory of relativity.

Using Bell's equation 2 on local hidden variables, this study was able to predict the observable quantities in Bell test on correlated photons and compare them with the experimental data. The experimental data here are widely known in the scientific communities and in the general public. Section 2 provides some perspectives on wave mechanics. This overview includes: quantum wavefunction, classical optics and photon polarization. Section 3 presents Bell's incompatibility criterion (local hidden variables vs quantum mechanics) and local (photon polarization) variables formulation. Section 4 provides the hidden variables predictions and compares them with the quantum mechanics. Section 5 compares the Bell inequalities (local hidden variables, quantum mechanics and Bell experiments).

## 2. Perspectives on Wave Mechanics

*On quantum wavefunction,* Schrödinger began with the Hamilton-Jacobi equation and laid out the foundation of quantum wave mechanics [8-11]. This is what turns wavefunction (Ψ) into a continuous quantity over all space, completes Hamilton's optical mechanical analogy and at the same time quantities the system. In quantum mechanics, the expected value for a physical quantity is obtained by multiplying the corresponding operator by the wavefunction Ψ and its complex conjugate Ψ* as for momentum $\langle p \rangle = \int \Psi^* (\hat{p}) \Psi \, dx$, position $\langle x \rangle = \int \Psi^* (\hat{x}) \Psi \, dx$ and likewise for other quantities; here, $\hat{p}$ and $\hat{x}$ are the momentum and position operators, respectively. These two non-commuting quantities can not be simultaneously real.

EPR first showed that when two particles emitted from the same source can be entangled by two orthogonal wavefunctions (superposition), subsequently, two non-commuting variables can be simultaneously real. As it violates the special theory of relativity (no information can travel faster than the speed of light), EPR argued that the quantum mechanical description given by the wavefunction was incomplete.

One concluding remark is that the Schrödinger wave equation was derived (engineered) by satisfying several known conditions of a quantum system (i.e., Planck-Einstein postulate, de Brogie hypothesis, wavefunctions linear superposition), not on the same character of a theory about a natural law, for example energy quanta in the black body radiation by Planck or principle of covariance in the relativity by Einstein. An argument here is that quantum wave mechanics will never be able to provide a complete picture of a physical system.

*On Classical Optics,* there has been extensive research on optical entanglement and violation of the Bell inequalities. On quantum technology, Heinrich *et al*. [12] provided a review on extended research on the fundamental principles and practical applications of quantum coherence in nanoscale systems. The authors emphasized the importance of controlling matter and waves at the nanometer scale to increase quantification energies, interaction strengths and coherence times and unlock novel functionalities explicitly enabled by quantum mechanics.

On wave-particle duality, a tentative theory of light quanta was first given by de Broglie [23] and a few years later the experimental data on the scattering of electrons by a single crystal of nickel was presented by Davisson and Germer [24]. Orefice *et al.* [25] provided an informative review of mathematical development, starting from de Broglie's and Schrödinger's foundation of wave mechanics, on the basis of numerical computations of the dynamics of wave-particle duality. On quantum reality, the author Anil Ananthaswamy provided a succinct summary of the status of today's experiments on Thomas Young double-slit interference patterns [13]. This summary included several well known papers on optical experiments [14-22]. For the moment, we are just appreciating the perspective provided by these articles on the modern tools and issues that address some intrinsic nature of particles.

*On photon interaction,* a photon will pass through a filter if its electric field is parallel to the filter and will be blocked by a filter oriented perpendicularly. For any other orientation angle, θ, the expected value for the photon will pass is cos θ. The intensity of the transmitted light varies as the square of the cosine of the angle as Malus's law. The photon polarization states will change depending on the orientation of the filter. Such a change in photon states can easily be viewed if photons interact with 2 or 3 separate filters at different orientations. According to the EPR picture of physical reality, each uninterrupted photon along its flight path has a specific polarization state (which is given or altered by the equipment) that can be known without disturbing it. Bell provided an interesting illustration of a spin half particle in a pure spin state vector relative to the observation filter and of the uncertainty of the desired expected value for individual instances. Let us now imagine the polarizing states of all photon pairs produced from the Aspect *et al.* experiment: a (J=0)-(J=1)-(J=0) cascade in calcium-40 excited by two-photon absorption, with the use of two single-mode lasers. Aspect *et al.* was able to produce pairs of visible photons (at wavelengths of 551. 3 *nm* and 422. 7 *nm*) at a typical rate of $5x10^7$ $s^{-1}$. The flux of photon pairs in the experiment was safely assumed to be an ensemble of all polarization states over the entire $360^0$ hemisphere. The quantities of Bell inequalities were subsequently measured, and violations of Bell inequalities were confirmed.

# 3. Hidden Variables Formulation

Until now, many Bell test experiments were conducted on correlated photons and came to conclusion in favor of quantum mechanics. None of these experimental studies was able to show any local variable predictions. This section now presents local causality equations (photon polarization as local hidden variables) for four observable quantities in Bell test on correlated photons. Follow up sections then provide systematic hidden variables predictions and compare them with quantum mechanics and Bell test experiment.

## *3.1. Bell Inequalities*

Clauser *et al.* [2] presented a generalization of Bell inequalities by avoiding Bell's experimentally unrealistic restriction for some photon pairs with perfect correlation. Bell tests by Aspect *et al.* [7] on pairs of visible photons emitted in atomic radiative cascades involve the transposition of Einstein-Podolsky-Rosen-Bohm *gedanken* experiment (Figure 1).

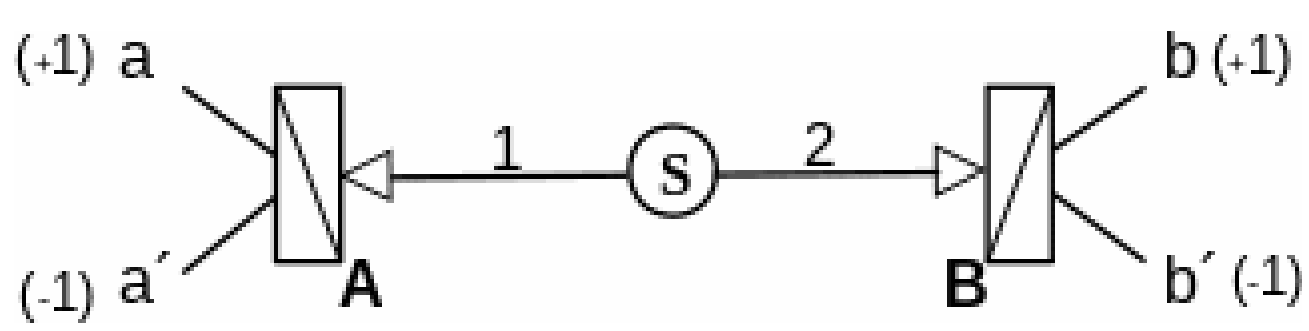

**Figure l.** Experimental setup similar to the Einstein-Podolsky-Rosen-Bohm *gedanken* experiment. Two polarizers in two different orientations: a and a' on side *A* and b and b' on side *B*. Pairs of photons from a source in a non factorizing state (similar to two spin-1/2 particles in a singlet state).

John Bell showed that the expected value of two independent measurements, $P(\vec{a},\vec{b})=\int A(\vec{a},\lambda)B(\vec{b},\lambda)\rho(\lambda)d\lambda$ (Bell's equation 2), is incompatible with quantum mechanical expectation value, $\langle\vec{\sigma}_1.\vec{a},\vec{\sigma}_2.\vec{b}\rangle=-\vec{a}.\vec{b}$ (Bell's equation 3). Here, $\vec{a},\vec{b}$ are the orientation vectors of the polarizers, $\vec{\lambda}$ is a unit vector representing the photon polarization states. Bell described the parameter $\vec{\lambda}$, where it stands for any number of variables and the dependencies of *A* and *B* polarizers, or it would have dynamical significance and laws of motion one would envision. By assuming a normalized density function, $\int \rho(\lambda)\, d\lambda = 1$, of an ensemble of the state vectors over the entire $360^0$ hemisphere, the generalized Bell inequality was given as:

$$-2 \leq S \leq 2 \qquad (1)$$

where $S = E(\vec{a}, \vec{b}) - E(\vec{a}, \vec{b'}) + E(\vec{a'}, \vec{b}) + E(\vec{a'}, \vec{b'})$. The four quantities in the inequality are the expected values of photons at two observation sites *(A, B)* as shown in Figure 1. Let us use the symbol *"p"* when a particle passes and the symbol *"n"* when blocked. Based on the expected values, first quantity is then expressed as: $E(\vec{a}, \vec{b}) = P_{pp}(\vec{a}, \vec{b}) + P_{nn}(\vec{a}, \vec{b}) - P_{pn}(\vec{a}, \vec{b}) - P_{np}(\vec{a}, \vec{b})$ and, similarly, the other 3 quantities.

An argument on statistical independence arose this very important question: is this normalized density function, $\int \rho(\lambda)\, d\lambda = 1$, representing the entire population (polarization states over $360^0$) or any individual instances (if it exists)? *This study discovered that the Bell inequality (equation 1) on normalized density function holds for photon polarization at individual instances but it does not hold over the entire population in a given system.* Two independent measurements using local hidden variables are not entirely incompatible with the quantum mechanical expectation. This will be clear in the next section when we analyze the hidden variables predictions over the entire $360^0$ hemisphere and for individual instances.

Mermin [26] provided some interesting perspectives on hidden variables and the Bell theorem. Batle and Casas [27] computed the density matrix for particles of spin- 1/2 transverse magnetic field and the Bell inequality. They analyzed the entanglement and nonlocality through the maximal violation of a Bell inequality. Scully and Zubairy [28] presented a review on the EPR arguments concerning the incompleteness of quantum mechanics and Bell inequality violation by quantum mechanics. They claimed that a nonlocal hidden variables theory can be developed which is in agreement with quantum theory.

## *3.2. Hidden Variables*

Suppose we are conducting Bell tests on correlated photon pairs emitted in atomic cascades analogous to Aspect *et al.* (Figure 1). Let us define the polarization states of *N* emitted pairs of photons over the entire hemisphere by $N$-unit vector, $\left(\vec{\lambda}_k, k=1..N\right)$ and $N$ independent orientations of each polarizer at A and B relative to the photon polarization axis, $\left(\vec{a}_i, i=1..N, \vec{b}_j, j=1..N\right)$. Using Bell's equation 2, the expected value of two independent measurements can be expressed as:

$$P(\vec{a},\vec{b})=\left(\int A(\vec{a},\lambda)B(\vec{b},\lambda)\rho(\lambda)d\lambda\right)=\sum_{k=1}^{N} A\left(\vec{a}_i,\vec{\lambda}_k\right)B\left(\vec{b}_j,\vec{\lambda}_k\right) \quad (2)$$

Computationally, we do not have any difficulties in predicting this probability field, $P\left(\vec{a}_i,\vec{b}_j\right)$ for any given photon pairs whence ($\left(\vec{a}_i,\vec{\lambda}_k\right)$) and ($\left(\vec{b}_j,\vec{\lambda}_k\right)$) are the corresponding angles between the polarizer orientation and photon state vector $\left(\vec{\lambda}_k\right)$. For such a setting, the expected values are $A\left(\vec{a}_i,\vec{\lambda}_k\right)=\cos\left(\vec{a}_i,\vec{\lambda}_k\right); B\left(\vec{b}_j,\vec{\lambda}_k\right)=\cos\left(\vec{b}_j,\vec{\lambda}_k\right)$ for photons that passed through filters $A$ and $B$ (similarly, by the *sine function* when the photon is blocked). Bell viewed this picture from an observation perspective and concluded, *"...the statistical features of quantum mechanics arise because the value of this variable is unknown in individual instances."*

For an ensemble of $N$ polarization states (over $360^0$ hemisphere), the four quantities in Bell test can be defined as:

$$\bar{P}^{\lambda}_{pp}\left(\vec{a}_i,\vec{b}_j\right)=\sum_{k=1}^{N}\cos\left(\vec{a}_i,\vec{\lambda}_k\right)\cos\left(\vec{b}_j,\vec{\lambda}_k\right) \quad (3a)$$

$$\bar{P}^{\lambda}_{nn}\left(\vec{a}_i,\vec{b}_j\right)=\sum_{k=1}^{N}\sin\left(\vec{a}_i,\vec{\lambda}_k\right)\sin\left(\vec{b}_j,\vec{\lambda}_k\right) \quad (3b)$$

$$\bar{P}^{\lambda}_{pn}\left(\vec{a}_i,\vec{b}_j\right)=\sum_{k=1}^{N}\cos\left(\vec{a}_i,\vec{\lambda}_k\right)\sin\left(\vec{b}_j,\vec{\lambda}_k\right) \quad (3c)$$

$$\bar{P}^{\lambda}_{np}\left(\vec{a}_i,\vec{b}_j\right)=\sum_{k=1}^{N}\sin\left(\vec{a}_i,\vec{\lambda}_k\right)\cos\left(\vec{b}_j,\vec{\lambda}_k\right) \quad (3d)$$

The first two equations (3a) and (3b) describe when both photons pass or are blocked, respectively. The last two equations (3c) and (3d) describe when one passes by filter $A$ and the other is blocked by filter $B$, respectively. Sign "*bar*" indicates an average (an ensemble of all polarization states).

A systematic prediction on correlated photon pairs will show next that these four equations in fact fully describe four quantities in Bell test and have a complete agreement with the experiments. This finding gives us great satisfaction as perceived naturalness of each element in the Bell test (photon-filter interaction did not violate the special theory of relativity).

### *3.3. Quantum Mechanics*

In quantum mechanics, two distance measurements are treated into a unified system. Bell's equation 3 describes the quantum mechanical expectation value for two spin-1/2 particles in a singlet state. The expectation values for all photons pairs in a unified can be defined as:

$$\bar{P}_{pp}^{QM}(\vec{a}_i, \vec{b}_j) = \frac{1}{2}\cos^2(\vec{a}_i, \vec{b}_j) \tag{4a}$$

$$\bar{P}_{nn}^{QM}(\vec{a}_i, \vec{b}_j) = \frac{1}{2}\sin^2(\vec{a}_i, \vec{b}_j) \tag{4b}$$

Equations 4a and 4b describe when both photons are passed or blocked, respectively. As we can see, quantum mechanics is unable to quantify four observable quantities in the Bell test. Our objection here is that the quantum mechanical description given by wavefunction is incomplete and it is incapable of providing a complete picture of a physical system (i.e. no intrinsic character). Einstein was not able to accept such a description given by the wavefunction (but unable to provide a concrete theoretical and experimental details).

## 4. Bell Test Prediction

This section presents the local hidden variables predictions for each quantity in the Bell test (equations 3a to 3d) and compared them with quantum mechanics (equations 4a and 4b). The results show that the hidden variables predictions are not entirely incompatible with quantum mechanics (which is a contradiction with Bell's argument).

### *4.1. Hidden Variables Prediction*

The hidden variables predictions were presented step by step in Appendix A. The following are three main steps in the hidden variables predictions on correlated pairs of photons:

*Step 1*, expectation values for each polarization state (individual instances), $P_{pp}^{\lambda_k}(\vec{a}_i,\vec{b}_j)$, $P_{nn}^{\lambda_k}(\vec{a}_i,\vec{b}_j)$, $P_{pn}^{\lambda_k}(\vec{a}_i,\vec{b}_j)$, $P_{np}^{\lambda_k}(\vec{a}_i,\vec{b}_j)$ (let set up $11.25^0$ equally spaced intervals $(i=1..32, j=1..32, k=1..32)$ over $360^0$ hemisphere that will sufficiently remove some disturbances caused by angles difference between polarizers orientation and photons polarization).

*Step 2*, expectation values (individual instances) for four observable quantities in Bell test, $E^{\lambda_k}(\vec{a}_i,\vec{b}_j)=P_{pp}^{\lambda_k}(\vec{a}_i,\vec{b}_j)+P_{nn}^{\lambda_k}(\vec{a}_i,\vec{b}_j)-P_{pn}^{\lambda_k}(\vec{a}_i,\vec{b}_j)-P_{np}^{\lambda_k}(\vec{a}_i,\vec{b}_j)$, followed by predictions of Bell inequalities for each polarization state, $S^{\lambda_k}=E^{\lambda_k}(\vec{a}_i,\vec{b}_i)-E^{\lambda_k}(\vec{a}_i,\vec{b'}_j)+E^{\lambda_k}(\vec{a'}_i,\vec{b}_j)+E^{\lambda_k}(\vec{a'}_i,\vec{b'}_j)$.

*Step 3*, mean expectation values for all polarization states over the entire hemisphere (population), $\bar{E}(\vec{a}_i,\vec{b}_j)=\bar{P}_{pp}(\vec{a}_i,\vec{b}_j)+\bar{P}_{nn}(\vec{a}_i,\vec{b}_j)-\bar{P}_{pn}(\vec{a}_i,\vec{b}_j)-\bar{P}_{np}(\vec{a}_i,\vec{b}_j)$, followed by predictions of Bell inequalities for the test system, $\bar{S}=\bar{E}(\vec{a},\vec{b})-\bar{E}(\vec{a},\vec{b'})+\bar{E}(\vec{a'},\vec{b})+\bar{E}(\vec{a'},\vec{b'})$.

For clarity (such as index *(i,j,k)*, photon interaction angle $(a_i,\lambda_k)$), the following is a sample calculation for photons within a visible spectrum: $\cos(a_3,\lambda_2)\cos(b_3,\lambda_2) = \cos(2\pi\cos(a_3-\lambda_2)/180)*\cos(2\pi\cos(b_3-\lambda_2)/180) = 0.854$ when $\lambda_2 = 11.25^0$ (correlated photons), $a_3=22.5^0$, $b_3=22.5^0$ (filter orientations).

## *4.2. Hidden Variables vs Quantum Mechanics*

Let us compare the hidden variables prediction with the quantum mechanics when both photons passed (equation 3a vs equation 4a). These two predictions, $\bar{P}_{pp}^{\lambda}(\vec{a}_i,\vec{b}_j)\equiv\bar{P}_{pp}^{QM}(\vec{a}_i,\vec{b}_j)$, are nearly identical**.** Figure 2 showed a sample comparison for five selected orientations of polarizer *B* $(\vec{b}_1,\vec{b}_3,\vec{b}_5,\vec{b}_7,\vec{b}_9)$ and all 32 orientations of polarizer *A*. When photons passed one filter and block by another, the results showed very similar agreement with quantum mechanics. Interestingly, there are some well defined  disagreements (small to large periodically). This disagreement is not the same as Bell's incompatibility argument between local hidden variables (Bell's equation 2) and quantum mechanics (Bell's equation 3).

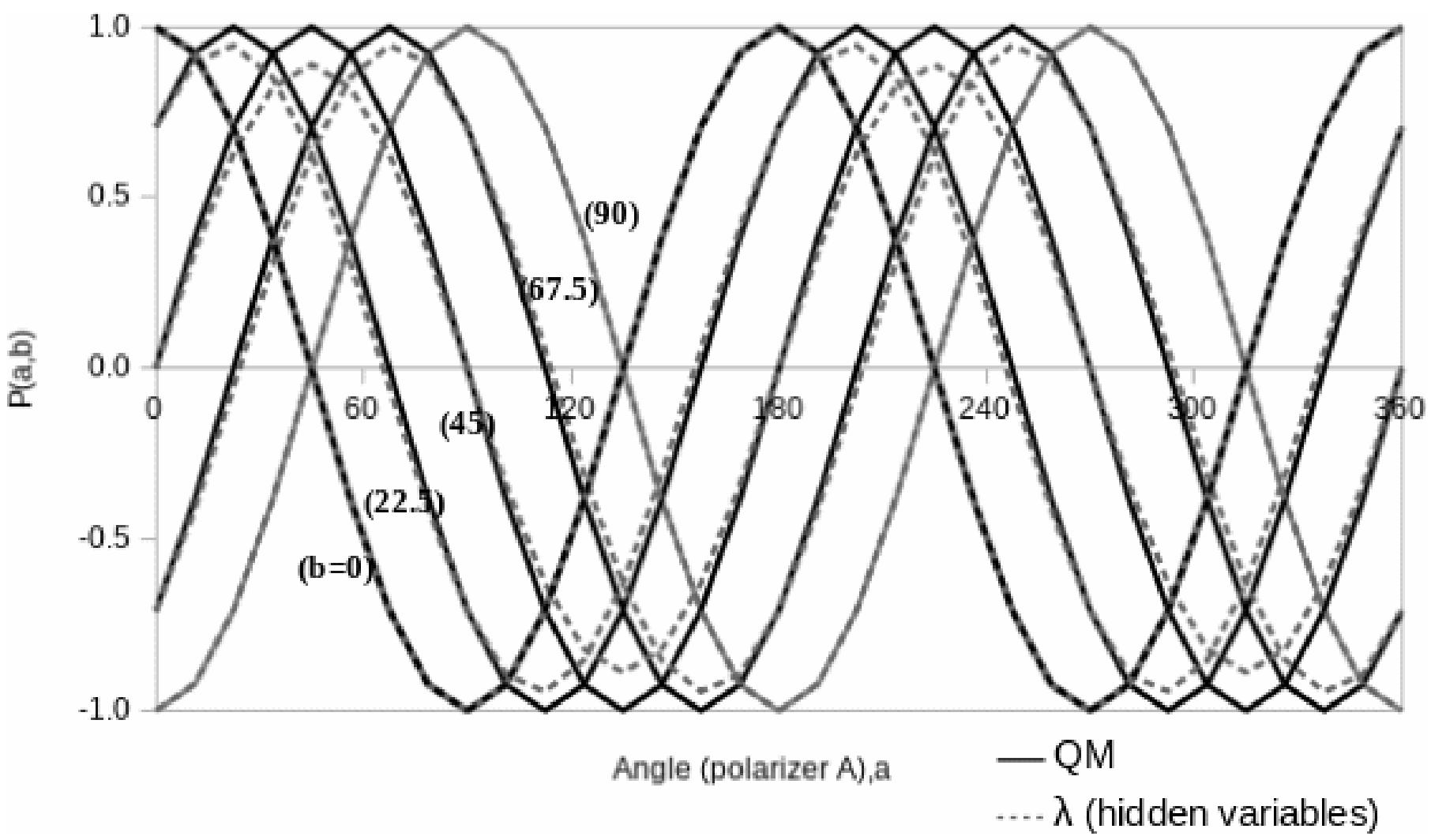

**Figure 2.** Expectation values when photons passed both filters - local hidden variables (equation 3a) vs quantum mechanics (equation 4a).

This overall agreement (local hidden variables vs quantum mechanics) is a contradiction with Bell's argument that the hidden variables prediction is incompatible with quantum mechanics. Next, we will get concrete proof that the local hidden variables (photon polarization) prediction given by the equations 3a to 3d is correct (physical) and the quantum mechanics prediction given by equations 4a and 4b is a system approximation.

## 5. Bell Inequalities Prediction

Until now, it is an overwhelming consensus that any local hidden variables predictions will always be bounded by this limit $-2 \leq S \leq 2$ and quantum mechanics can violate this inequality. The results here will show that the Bell inequality holds only for photon polarization at individual instances but it does not hold over the entire population in a given system. *The Bell inequality is not an incompatibility criterion between local hidden variables vs quantum mechanics rather it is a criterion for an individual nature vs population dynamics in a given system*.

*5.1. Individual Polarization*

O*n Individual Instances*, let us analyze a few details of the expectation values for photon polarization at individual instances ($\vec{\lambda}_k, k=1..32$). Figure 3 showed a sample of the expected values for several individual instances $\left(\vec{\lambda}_1=0, \vec{\lambda}_2=11.25, \vec{\lambda}_3=22.5, \vec{\lambda}_4=33.75\right)$ when the polarizer *B* orientation vector is set to $\vec{b}_1=0$ and polarizer *A* for all orientations ($\vec{a}_i, i=1..32$). This prediction for individual polarization state may support Einstein pictures of physical reality (observables by experiment). Such hidden variables predictions for individual polarization have no realization in quantum mechanics.

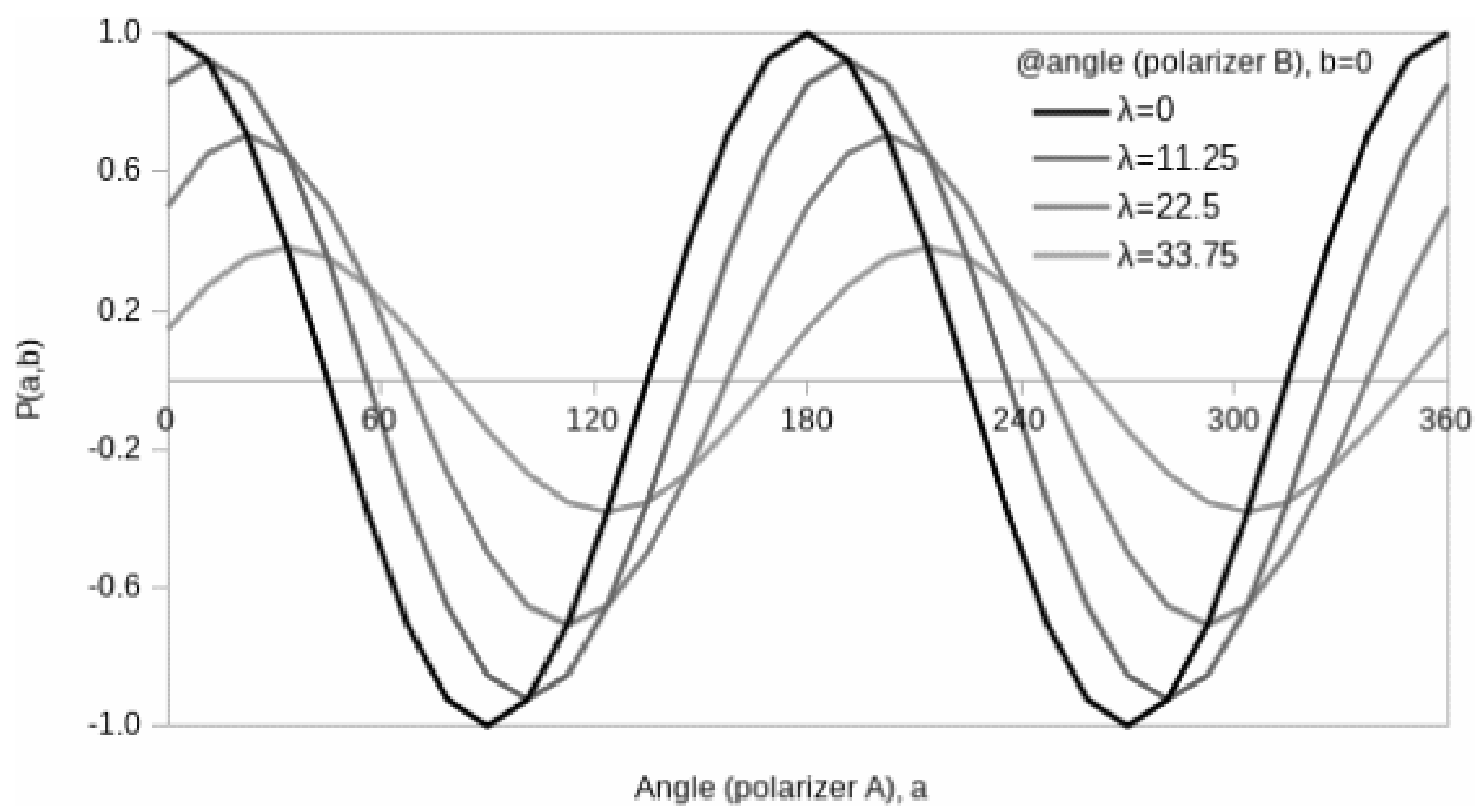

**Figure 3.** Hidden variables prediction (equation 3a) for photons passed both filters at four selected polarization states (intrinsic properties, no realization in quantum mechanics).

Table 1 summarizes the Bell inequalities for several polarizer settings at observation sites *A* and *B* (only the upper and lower values are shown). The local variables (photon polarization) predictions fully agreed with the Bell inequality ($-2 \le S \le 2$). The predicted Bell inequalities in Table 1 can be experimentally verified if we can somehow set up a test for photon pairs over a specified polarization spectrum (such as polarization states over a selected interval of $\theta=11.25$). Using the existing setup in Figure 1, such a test may be possible if we use an optical filter that lets photons of a specified polarization pass through while blocking photons of other polarization (before entering two distance measurements). The criterion in designing any such optical filter should be blocking all photons of unwanted polarization and allowing only photons of specific polarization (any disturbances need to be justified). Next section presents several well known Bell tests over the entire $360^0$ hemisphere and compares them with local variables (photon polarization) and quantum mechanics predictions.

**Table 1**

Bell inequalities for individual polarization states (shown only the limits) *(Table A.5 in appendix A included the predicted values) (the Bell inequality,-2≤S≤2, holds).*

| ***Polarizers setup*** ***θ=(a,b)=(b,a´)=(a´,b´)*** | *S* ***limit for each setup*** ***(32*** *S* ***values for 32 polarization states)*** |
|---|---|
| $\theta=11.25, 78.25$ | $0.541 \le S \le 1.848, -1.848 \le S \le -0.541$ |
| $\theta=22.5, 67.5$ | $S=1.414, S \le -1.414$ |
| $\theta=33.75, 56.25$ | $0.383 \le S \le 1.689, -1.689 \le S \le -0.383$ |
| $\theta=45, 135$ | $-1.0 \le S \le 1.0$ |
| $\theta=90, 180$ | $-2.0 \le S \le 2.0$ |

## *5.2. Polarization Ensembles*

*Over $360^0$ hemisphere (population)*, the predictions (Table 2) for several orientations of the polarizers showed a clear violation of the Bell inequalities ($-2 \le S \le 2$). Most surprisingly, the hidden variable predictions are in complete agreement with the quantum mechanics and Aspect *et al.* prediction on lab data. This is in contradiction with an affirmative conclusion in favor of quantum mechanics made by the published experimental studies (until now).

**Table 2**

Bell inequalities over the entire 360$^0$ hemisphere (polarization population) *(Table A.7 in appendix A showed more details) (the Bell inequality, -2≤S≤2, does not holds in all three predictions:* local hidden variables, quantum mechanics (QM), lab data*).*

| Test | 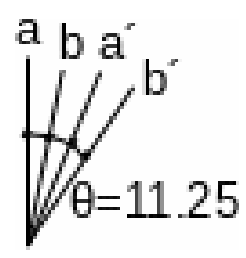 | 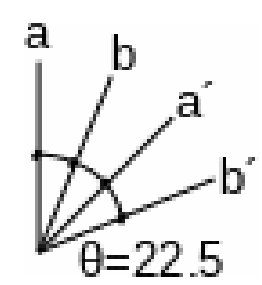 | 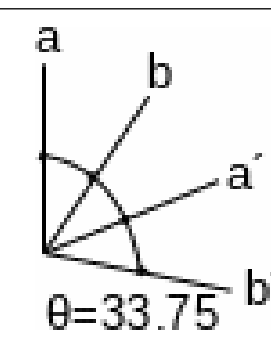 |
|---|---|---|---|
| λ (hidden variables)[§] | 2.328 | 2.794 | 2.047 |
| QM | 2.389 | 2.828 | 2.072 |
| Aspect *et al.* Exp. | - | 2.697±0.015 | - |

[§]*hidden variables (photon polarization) prediction in this study (previously not known).*

Table 3 showed the Bell test analysis for two independent experiments (relatively large scale setup). The hidden variables predictions here are consistent with the experimental quantities and have a complete agreement. In contrast, the following measured quantities in Bell test have no realization in a unified quantum mechanical description: E(a,b), E(a,b′), E(a′,b) and E(a′,b′). This clearly demonstrates that the local hidden variables (photon polarization states) prediction is correct (physical) and the quantum mechanics prediction on a unified system is a system approximation.

**Table 3**

Bell test analysis over the entire 360$^0$ hemisphere (polarization population) *(the Bell inequality, -2≤S≤2, does not holds in all three predictions:* local hidden variables, quantum mechanics (QM), lab data*).*

| 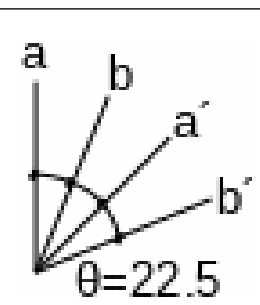 | ***E(a,b)*** | ***E(a,b′)*** | ***E(a′,b)*** | ***E(a′,b′)*** | ***S*** |
|---|---|---|---|---|---|
| λ (hidden variables)[§] | 0.686 | -0.728 | 0.669 | 0.711 | 2.794 |
| QM | - | - | - | - | 2.828 |
| Cosmic Bell Test | 0.670 | -0.739 | 0.637 | 0.628 | 2.674 |
| Laser Entangled Photons | 0.559 | -0.591 | 0.560 | 0.820 | 2.530 |

[§]*hidden variables (photon polarization) prediction in this study (previously not known).*

The Cosmic Bell Test was conducted by a team of Anton Zeilinger. This is a large version of Clauser and Freedman's Bell test experiment. In this Cosmic Bell test, the source of the entangled particles was about a third of a mile from each of the detectors. The team sent perfectly timed pairs of photons through the air to each detector. At the same time, the telescopes collect light from two extremely far-off, extremely bright galaxies called quasars. Random variations in this light controlled which filters were used to measure the photon pairs. In such controlling, it was incredibly unlikely that anything could be influencing the random nature of the filters. The Laser Entangled Photons Bell test was presented in a documentary on quantum physics by Jim Al-Khalili (very informative for the scientific communities and for the general public). This is a modern version of an experiment first carried out by John Clauser and then Alain Aspect. In this experiment, a crystal converts laser light into pairs of entangled light quanta, photons, making two very precise beams. These correlated photons then passed through the detectors.

### *5.3. Summary Results*

Table 1 presents Bell inequalities (local hidden variables prediction) for photon polarization at individual instances. The predictions are bounded by the Bell inequality, $-2 \leq S \leq 2$. Table 2 presents the Bell inequalities for the entire $360^0$ hemisphere. The local variables (polarization) predictions have a complete agreement with the experiments and quantum mechanics. All three predictions (local hidden variables, quantum mechanics, lab data) over polarization ensembles showed a clear violation of the Bell inequality. The results do not agree with the following conclusion made by the earlier studies: *quantum mechanics can not be replaced by a theory that uses hidden variables.* Table 3 shows all four observable quantities in Bell test and a complete agreement between local hidden variables (polarization) and experiments. This demonstrates that the photon-filter interaction at a distance was a local cause (no violation of special theory of relativity).

The Bell inequality was derived assuming normalized density function for the correlated photon pairs or spin half particles by Stern-Gerlach magnets. As a result, the Bell inequality is unable to distinguish any individual photon polarization states from its population in a given experiment. Considering statistical independence, one may assume these two systems can be analogous: polarization ensembles on normalized density function vs polarization at individual instances. This suggests that violation of the Bell inequality may likely occur in many classical optical settings or even in large scale molecular dynamics. A perspective on emerging properties (individual instances vs population) in particles and biological systems is given in appendix B.

## 6. Conclusion

This study revealed that the Bell inequality holds for individual instances but it does not hold over the entire population in the Bell tests on photon polarization. The Bell inequality is not an incompatibility criterion between local hidden variables vs quantum mechanics rather it is a criterion for an individual nature vs population dynamics. For polarization over the hemisphere, violation of the Bell inequality occurred in all three predictions: local hidden variables, quantum mechanics and experiment. For polarization at individual instances, the hidden variables predictions were bounded by the Bell inequality. The study recommends a possible Bell test on photon pairs for a specified spectrum of polarization states by utilizing the existing Bell set up (Figure 1). Most importantly, this study arrived at the following conclusion with great satisfaction (as perceived naturalness of each element in the Bell test): the local hidden variables (photon polarization) prediction is correct (real) and the quantum mechanics description given by wavefunction in a unified system is a system approximation. The interaction of photons at a distance was a local cause (no violation of the special theory of relativity).

## Acknowledgments

The author would like to thank D. Coombe, M. Barton, R. Palfrey for insightful discussions. This work was supported by Alberta Computational Biochemistry Lab.

## Declaration of interest Statement

The author declares no known competing financial interests or personal relationships that could have appeared to influence the work reported in this paper.

## Data Availability

The author confirms that all data generated or analysed during this study are included in this published article.

# Appendix A : Local Hidden Variables Prediction

## A.1 Hidden Variables Equation

### *A.1.1 Photon Filter Interaction*

Equations A-1 and A-2 describe when both photons pass (subscript “*pp*”) or are blocked (subscript “*nn*”), respectively. Equations (A-3) and (A-4) describe when one passes by filter A and the other is blocked by filter B (subscript “*pn*” *or* “*np*”), respectively. The indexes *i*, *j* represent the filter orientations, and *k* for photon polarization state.

$$P_{pp}^{\lambda_k}\left(\vec{a}_i,\vec{b}_j\right)=\cos\left(\vec{a}_i,\vec{\lambda}_k\right)\cos\left(\vec{b}_j,\vec{\lambda}_k\right) \qquad \text{(A-1)}$$

$$P_{nn}^{\lambda_k}\left(\vec{a}_i,\vec{b}_j\right)=\sin\left(\vec{a}_i,\vec{\lambda}_k\right)\sin\left(\vec{b}_j,\vec{\lambda}_k\right) \qquad \text{(A-2)}$$

$$P_{pn}^{\lambda_k}\left(\vec{a}_i,\vec{b}_j\right)=\cos\left(\vec{a}_i,\vec{\lambda}_k\right)\sin\left(\vec{b}_j,\vec{\lambda}_k\right) \qquad \text{(A-3)}$$

$$P_{np}^{\lambda_k}\left(\vec{a}_i,\vec{b}_j\right)=\sin\left(\vec{a}_i,\vec{\lambda}_k\right)\cos\left(\vec{b}_j,\vec{\lambda}_k\right) \qquad \text{(A-4)}$$

### *A.1.2 Bell Ineqaulities - Individual Polarization*

For experimental setup (Figure 1 in main body), the bell inequalities was given as:

$$S^{\lambda_k}=E^{\lambda_k}\left(\vec{a}_i,\vec{b}_i\right)-E^{\lambda_k}\left(\vec{a}_i,\vec{b}'_j\right)+E^{\lambda_k}\left(\vec{a}'_i,\vec{b}_j\right)+E^{\lambda_k}\left(\vec{a}'_i,\vec{b}'_j\right) \qquad \text{(A-5a)}$$

where the expected value was defined as:

$$E^{\lambda_k}\left(\vec{a}_i,\vec{b}_j\right)=P_{pp}^{\lambda_k}\left(\vec{a}_i,\vec{b}_j\right)+P_{nn}^{\lambda_k}\left(\vec{a}_i,\vec{b}_j\right)-P_{pn}^{\lambda_k}\left(\vec{a}_i,\vec{b}_j\right)-P_{np}^{\lambda_k}\left(\vec{a}_i,\vec{b}_j\right) \qquad \text{(A-5b)}$$

### *A.1.3 Bell Inequalities - Entire Population*

The bell inequalities for the entire population (using all polarization states over $360^0$ hemisphere) was defined as:

$$\bar{S} = \bar{E}\left(\vec{a}, \vec{b}\right) - \bar{E}\left(\vec{a}, \vec{b'}\right) + \bar{E}\left(\vec{a'}, \vec{b}\right) + \bar{E}\left(\vec{a'}, \vec{b'}\right) \qquad \text{(A-6a)}$$

where the expected value is an average over all polarization states as:

$$\bar{E}\left(\vec{a}_i, \vec{b}_j\right) = \bar{P}_{pp}\left(\vec{a}_i, \vec{b}_j\right) + \bar{P}_{nn}\left(\vec{a}_i, \vec{b}_j\right) - \bar{P}_{pn}\left(\vec{a}_i, \vec{b}_j\right) - \bar{P}_{np}\left(\vec{a}_i, \vec{b}_j\right) \qquad \text{(A-6b)}$$

# A.2 Bell Test Prediction

## *A.2.1 Photon Filter Interaction*

Tables A.1 to A.4 showed the photon interaction predictions using the equation A-1 to A-4, respectively. This approach was applied for all polarization states *(k)* and filter orientations *(i, j)* over $360^0$ (evenly spaced). Note: predictions here are given for one polarization state at $\lambda_2=11.25^0$ and can be similarly predicted for other polarization (also not shown for filter orientations $a_{10}...a_{31}$ and $b_{10}...b_{31}$).

**Table A.1**

Prediction, $P_{pp}^{\lambda_k}(\vec{a}_i, \vec{b}_j)$, for polarization state at $\lambda_2=11.25^0$.

| $a_i/b_j$ | $b_1$ (0) | $b_2$ (11.25) | $b_3$ (22.5) | $b_4$ (33.75) | $b_5$ (45) | $b_6$ (56.25) | $b_7$ (67.5) | $b_8$ (78.75) | $b_9$ (90) | .. | $b_{32}$ (348.75) |
|---|---|---|---|---|---|---|---|---|---|---|---|
| $\lambda_2$=11.25 | | | | | | | | | | | |
| $a_1$ (0) | 0.854 | 0.924 | 0.854 | 0.653 | 0.354 | 0.0 | -0.354 | -0.653 | -0.854 | | 0.653 |
| $a_2$ (11.25) | 0.924 | 1 | 0.924 | 0.707 | 0.383 | 0.0 | -0.383 | -0.707 | -0.924 | | 0.707 |
| $a_3$ (22.5) | 0.854 | 0.924 | 0.854 | 0.653 | 0.354 | 0.0 | -0.354 | -0.653 | -0.854 | | 0.653 |
| $a_4$ (33.75) | 0.653 | 0.707 | 0.653 | 0.5 | 0.271 | 0.0 | -0.271 | -0.5 | -0.653 | | 0.5 |
| $a_5$ (45) | 0.354 | 0.383 | 0.354 | 0.271 | 0.146 | 0.0 | -0.146 | -0.271 | -0.354 | | 0.271 |
| $a_6$ (56.25) | 0.0 | 0.0 | 0.0 | 0.0 | 0.0 | 0.0 | 0.0 | 0.0 | 0.0 | | 0.0 |
| $a_7$ (67.5) | -0.354 | -0.383 | -0.354 | -0.271 | -0.146 | 0.0 | 0.146 | 0.271 | 0.354 | | -0.271 |
| $a_8$ (78.75) | -0.653 | -0.707 | -0.653 | -0.5 | -0.271 | 0.0 | 0.271 | 0.5 | 0.653 | | -0.5 |
| $a_9$ (90) | -0.854 | -0.924 | -0.854 | -0.653 | -0.354 | 0.0 | 0.354 | 0.653 | 0.854 | | -0.653 |
| .. | | | | | | | | | | | |
| $a_{32}$ (348.75) | 0.653 | 0.707 | 0.653 | 0.5 | 0.271 | 0.0 | -0.271 | -0.5 | -0.653 | | 0.5 |

*For clarity, this is a sample calculation:* $\cos(a_3,\lambda_2)\cos(b_3,\lambda_2) = \cos(2\pi\cos(a_3-\lambda_2)/180)*\cos(2\pi\cos(b_3-\lambda_2)/180) = 0.854$ *when* $\lambda_1= 11.25^0$ *(correlated photons),* $a_3=22.5$, $b_3=22.5$.

**Table A.2**

Prediction, $P_{nn}^{\lambda_k}\left(\vec{a}_i, \vec{b}_j\right)$, for polarization state at $\lambda_2 = 11.25^0$.

| $a_i/b_j$ | $b_1$ (0) | $b_2$ (11.25) | $b_3$ (22.5) | $b_4$ (33.75) | $b_5$ (45) | $b_6$ (56.25) | $b_7$ (67.5) | $b_8$ (78.75) | $b_9$ (90) | .. | $b_{32}$ (348.75) |
|---|---|---|---|---|---|---|---|---|---|---|---|
| $\lambda_2$=11.25 | | | | | | | | | | | |
| $a_1$ (0) | 0.146 | 0.0 | -0.146 | -0.271 | -0.354 | -0.383 | -0.354 | -0.271 | -0.146 | | 0.271 |
| $a_2$ (11.25) | 0.0 | 0.0 | 0.0 | 0.0 | 0.0 | 0.0 | 0.0 | 0.0 | 0.0 | | 0.0 |
| $a_3$ (22.5) | -0.146 | 0.0 | 0.146 | 0.271 | 0.354 | 0.383 | 0.354 | 0.271 | 0.146 | | -0.271 |
| $a_4$ (33.75) | -0.271 | 0.0 | 0.271 | 0.5 | 0.653 | 0.707 | 0.653 | 0.5 | 0.271 | | -0.5 |
| $a_5$ (45) | -0.354 | 0.0 | 0.354 | 0.653 | 0.854 | 0.924 | 0.854 | 0.653 | 0.354 | | -0.653 |
| $a_6$ (56.25) | -0.383 | 0.0 | 0.383 | 0.707 | 0.924 | 1 | 0.924 | 0.707 | 0.383 | | -0.707 |
| $a_7$ (67.5) | -0.354 | 0.0 | 0.354 | 0.653 | 0.854 | 0.924 | 0.854 | 0.653 | 0.354 | | -0.653 |
| $a_8$ (78.75) | -0.271 | 0.0 | 0.271 | 0.5 | 0.653 | 0.707 | 0.653 | 0.5 | 0.271 | | -0.5 |
| $a_9$ (90) | -0.146 | 0.0 | 0.146 | 0.271 | 0.354 | 0.383 | 0.354 | 0.271 | 0.146 | | -0.271 |
| .. | | | | | | | | | | | |
| $a_{32}$ (348.75) | 0.271 | 0.0 | -0.271 | -0.5 | -0.653 | -0.707 | -0.653 | -0.5 | -0.271 | | 0.5 |

**Table A.3**

Prediction, $P_{pn}^{\lambda_k}\left(\vec{a}_i, \vec{b}_j\right)$, for polarization state at $\lambda_2 = 11.25^0$.

| $a_i/b_j$ | $b_1$ (0) | $b_2$ (11.25) | $b_3$ (22.5) | $b_4$ (33.75) | $b_5$ (45) | $b_6$ (56.25) | $b_7$ (67.5) | $b_8$ (78.75) | $b_9$ (90) | .. | $b_{32}$ (348.75) |
|---|---|---|---|---|---|---|---|---|---|---|---|
| $\lambda_2$=11.25 | | | | | | | | | | | |
| $a_1$ (0) | -0.354 | 0.0 | 0.354 | 0.653 | 0.854 | 0.924 | 0.854 | 0.653 | 0.354 | | -0.653 |
| $a_2$ (11.25) | -0.383 | 0.0 | 0.383 | 0.707 | 0.924 | 1 | 0.924 | 0.707 | 0.383 | | -0.707 |
| $a_3$ (22.5) | -0.354 | 0.0 | 0.354 | 0.653 | 0.854 | 0.924 | 0.854 | 0.653 | 0.354 | | -0.653 |
| $a_4$ (33.75) | -0.271 | 0.0 | 0.271 | 0.5 | 0.653 | 0.707 | 0.653 | 0.5 | 0.271 | | -0.5 |
| $a_5$ (45) | -0.146 | 0.0 | 0.146 | 0.271 | 0.354 | 0.383 | 0.354 | 0.271 | 0.146 | | -0.271 |
| $a_6$ (56.25) | 0.0 | 0.0 | 0.0 | 0.0 | 0.0 | 0.0 | 0.0 | 0.0 | 0.0 | | 0.0 |
| $a_7$ (67.5) | 0.146 | 0.0 | -0.146 | -0.271 | -0.354 | -0.383 | -0.354 | -0.271 | -0.146 | | 0.271 |
| $a_8$ (78.75) | 0.271 | 0.0 | -0.271 | -0.5 | -0.653 | -0.707 | -0.653 | -0.5 | -0.271 | | 0.5 |
| $a_9$ (90) | 0.354 | 0.0 | -0.354 | -0.653 | -0.854 | -0.924 | -0.854 | -0.653 | -0.354 | | 0.653 |
| .. | | | | | | | | | | | |
| $a_{32}$ (348.75) | -0.271 | 0.0 | 0.271 | 0.5 | 0.653 | 0.707 | 0.653 | 0.5 | 0.271 | | -0.5 |

**Table A.4**

Prediction, $P_{np}^{\lambda_k}\left(\vec{a}_i,\vec{b}_j\right)$, for polarization state at $\lambda_2=11.25^0$.

| $a_i/b_j$ | $b_1$ (0) | $b_2$ (11.25) | $b_3$ (22.5) | $b_4$ (33.75) | $b_5$ (45) | $b_6$ (56.25) | $b_7$ (67.5) | $b_8$ (78.75) | $b_9$ (90) | .. | $b_{32}$ (348.75) |
|---|---|---|---|---|---|---|---|---|---|---|---|
| $\lambda_2$=11.25 | | | | | | | | | | | |
| $a_1$ (0) | -0.354 | -0.383 | -0.354 | -0.271 | -0.146 | 0.0 | 0.146 | 0.271 | 0.354 | | -0.271 |
| $a_2$ (11.25) | 0.0 | 0.0 | 0.0 | 0.0 | 0.0 | 0.0 | 0.0 | 0.0 | 0.0 | | 0.0 |
| $a_3$ (22.5) | 0.354 | 0.383 | 0.354 | 0.271 | 0.146 | 0.0 | -0.146 | -0.271 | -0.354 | | 0.271 |
| $a_4$ (33.75) | 0.653 | 0.707 | 0.653 | 0.5 | 0.271 | 0.0 | -0.271 | -0.5 | -0.653 | | 0.5 |
| $a_5$ (45) | 0.854 | 0.924 | 0.854 | 0.653 | 0.354 | 0.0 | -0.354 | -0.653 | -0.854 | | 0.653 |
| $a_6$ (56.25) | 0.924 | 1 | 0.924 | 0.707 | 0.383 | 0.0 | -0.383 | -0.707 | -0.924 | | 0.707 |
| $a_7$ (67.5) | 0.854 | 0.924 | 0.854 | 0.653 | 0.354 | 0.0 | -0.354 | -0.653 | -0.854 | | 0.653 |
| $a_8$ (78.75) | 0.653 | 0.707 | 0.653 | 0.5 | 0.271 | 0.0 | -0.271 | -0.5 | -0.653 | | 0.5 |
| $a_9$ (90) | 0.354 | 0.383 | 0.354 | 0.271 | 0.146 | 0.0 | -0.146 | -0.271 | -0.354 | | 0.271 |
| .. | | | | | | | | | | | |
| $a_{32}$ (348.75) | -0.653 | -0.707 | -0.653 | -0.5 | -0.271 | 0.0 | 0.271 | 0.5 | 0.653 | | -0.5 |

## *A.2.2 Bell Inequalities - Individual Polarization*

For all settings *(i,j,k)*, the expected values were predicted using equation A-5b. The Bell inequalities were then calculated for several selected cases. Table A.5 showed the expected values for $\lambda_2=11.25^0$ (sample prediction). Table A.6 summarized the Bell inequalities for individual polarization (discussed in main text Table 3).

**Table A.5**

Prediction, $E^{\lambda_k}\left(\vec{a}_i,\vec{b}_j\right)$, for polarization state at $\lambda_2=11.25^0$.

| $a_i/b_j$ | $b_1$ (0) | $b_2$ (11.25) | $b_3$ (22.5) | $b_4$ (33.75) | $b_5$ (45) | $b_6$ (56.25) | $b_7$ (67.5) | $b_8$ (78.75) | $b_9$ (90) | .. | $b_{32}$ (348.75) |
|---|---|---|---|---|---|---|---|---|---|---|---|
| $\lambda_2$=11.25 | | | | | | | | | | | |
| $a_1$ (0) | 0.854 | 0.653 | 0.354 | 0.0 | -0.354 | -0.653 | -0.854 | -0.924 | -0.854 | | 0.924 |
| $a_2$ (11.25) | 0.653 | 0.5 | 0.271 | 0.0 | -0.271 | -0.5 | -0.653 | -0.707 | -0.653 | | 0.707 |
| $a_3$ (22.5) | 0.354 | 0.271 | 0.146 | 0.0 | -0.146 | -0.271 | -0.354 | -0.383 | -0.354 | | 0.383 |
| $a_4$ (33.75) | 0.0 | 0.0 | 0.0 | 0.0 | 0.0 | 0.0 | 0.0 | 0.0 | 0.0 | | 0.0 |
| $a_5$ (45) | -0.354 | -0.271 | -0.146 | 0.0 | 0.146 | 0.271 | 0.354 | 0.383 | 0.354 | | -0.383 |
| $a_6$ (56.25) | -0.653 | -0.5 | -0.271 | 0.0 | 0.271 | 0.5 | 0.653 | 0.707 | 0.653 | | -0.707 |
| $a_7$ (67.5) | -0.854 | -0.653 | -0.354 | 0.0 | 0.354 | 0.653 | 0.854 | 0.924 | 0.854 | | -0.924 |
| $a_8$ (78.75) | -0.924 | -0.707 | -0.383 | 0.0 | 0.383 | 0.707 | 0.924 | 1 | 0.924 | | -1 |
| $a_9$ (90) | -0.854 | -0.653 | -0.354 | 0.0 | 0.354 | 0.653 | 0.854 | 0.924 | 0.854 | | -0.924 |
| .. | | | | | | | | | | | |
| $a_{32}$ (348.75) | 0.924 | 0.707 | 0.383 | 0.0 | -0.383 | -0.707 | -0.924 | -1 | -0.924 | | 1 |

**Table A.5**

Bell inequalities for individual polarization states for different test setting.

*(predicted Bell inequality,-2≤S≤2, holds).*

| 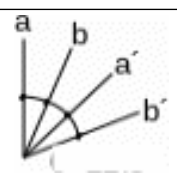 | $a_1b_2a_3b_4$ (11.25) | $a_1b_3a_5b_7$ (22.5) | $a_1b_4a_7b_{10}$ (33.75) | $a_1b_5a_9b_{13}$ (45) | $a_1b_6a_{11}b_{16}$ (56.25) | $a_1b_7a_{13}b_{19}$ (67.5) | $a_1b_8a_{15}b_{22}$ (78.75) | $a_1b_9a_{17}b_{25}$ (90) | $a_1b_{10}a_{19}b_{28}$ (101.25) | $A_1b_{11}a_{21}b_{31}$ (112.5) |
|---|---|---|---|---|---|---|---|---|---|---|
| **$\lambda_1$=0.0** | 0.5412 | 1.4142 | 0.7654 | -1.0000 | -1.3066 | -1.4142 | -1.8478 | -1.0000 | -0.5412 | -1.4142 |
| **$\lambda_2$=11.25** | 0.9239 | 1.4142 | 1.3066 | -0.7071 | -1.6892 | -1.4142 | -1.8478 | -1.7071 | -0.9239 | -1.4142 |
| **$\lambda_3$=22.5** | 1.4651 | 1.4142 | 1.6892 | 0.0000 | -1.6892 | -1.4142 | -1.4651 | -2.0000 | -1.4651 | -1.4142 |
| **$\lambda_4$=33.75** | 1.8478 | 1.4142 | 1.6892 | 0.7071 | -1.3066 | -1.4142 | -0.9239 | -1.7071 | -1.8478 | -1.4142 |
| **$\lambda_5$=45** | 1.8478 | 1.4142 | 1.3066 | 1.0000 | -0.7654 | -1.4142 | -0.5412 | -1.0000 | -1.8478 | -1.4142 |
| **$\lambda_6$=56.25** | 1.4651 | 1.4142 | 0.7654 | 0.7071 | -0.3827 | -1.4142 | -0.5412 | -0.2929 | -1.4651 | -1.4142 |
| **$\lambda_7$=67.5** | 0.9239 | 1.4142 | 0.3827 | 0.0000 | -0.3827 | -1.4142 | -0.9239 | 0.0000 | -0.9239 | -1.4142 |
| **$\lambda_8$=78.75** | 0.5412 | 1.4142 | 0.3827 | -0.7071 | -0.7654 | -1.4142 | -1.4651 | -0.2929 | -0.5412 | -1.4142 |
| **$\lambda_9$=90** | 0.5412 | 1.4142 | 0.7654 | -1.0000 | -1.3066 | -1.4142 | -1.8478 | -1.0000 | -0.5412 | -1.4142 |
| **$\lambda_{10}$=101.25** | 0.9239 | 1.4142 | 1.3066 | -0.7071 | -1.6892 | -1.4142 | -1.8478 | -1.7071 | -0.9239 | -1.4142 |
| **$\lambda_{11}$=112.5** | 1.4651 | 1.4142 | 1.6892 | 0.0000 | -1.6892 | -1.4142 | -1.4651 | -2.0000 | -1.4651 | -1.4142 |
| **$\lambda_{12}$=123.75** | 1.8478 | 1.4142 | 1.6892 | 0.7071 | -1.3066 | -1.4142 | -0.9239 | -1.7071 | -1.8478 | -1.4142 |
| **$\lambda_{13}$=135** | 1.8478 | 1.4142 | 1.3066 | 1.0000 | -0.7654 | -1.4142 | -0.5412 | -1.0000 | -1.8478 | -1.4142 |
| **$\lambda_{14}$=146.25** | 1.4650 | 1.4142 | 0.7655 | 0.7069 | -0.3829 | -1.4142 | -0.5414 | -0.2931 | -1.4650 | -1.4142 |
| **$\lambda_{15}$=157.5** | 0.9239 | 1.4142 | 0.3827 | 0.0000 | -0.3827 | -1.4142 | -0.9239 | 0.0000 | -0.9239 | -1.4142 |
| **$\lambda_{16}$=168.75** | 0.5412 | 1.4142 | 0.3827 | -0.7071 | -0.7654 | -1.4142 | -1.4651 | -0.2929 | -0.5412 | -1.4142 |
| **$\lambda_{17}$=180** | 0.5412 | 1.4142 | 0.7654 | -1.0000 | -1.3066 | -1.4142 | -1.8478 | -1.0000 | -0.5412 | -1.4142 |
| **$\lambda_{18}$=191.25** | 0.9239 | 1.4142 | 1.3066 | -0.7071 | -1.6892 | -1.4142 | -1.8478 | -1.7071 | -0.9239 | -1.4142 |
| **$\lambda_{19}$=202.5** | 1.4651 | 1.4142 | 1.6892 | 0.0024 | -1.6892 | -1.4142 | -1.4651 | -2.0000 | -1.4651 | -1.4142 |
| **$\lambda_{20}$=213.75** | 1.8478 | 1.4142 | 1.6892 | 0.7071 | -1.3066 | -1.4142 | -0.9239 | -1.7071 | -1.8478 | -1.4142 |
| **$\lambda_{21}$=225** | 1.8478 | 1.4142 | 1.3066 | 1.0000 | -0.7654 | -1.4142 | -0.5412 | -1.0000 | -1.8478 | -1.4142 |
| **$\lambda_{22}$=236.25** | 1.4651 | 1.4142 | 0.7654 | 0.7071 | -0.3827 | -1.4142 | -0.5412 | -0.2929 | -1.4651 | -1.4142 |
| **$\lambda_{23}$=247.5** | 0.9239 | 1.4142 | 0.3827 | 0.0000 | -0.3827 | -1.4142 | -0.9239 | 0.0000 | -0.9239 | -1.4142 |
| **$\lambda_{24}$=258.75** | 0.5412 | 1.4142 | 0.3827 | -0.7071 | -0.7654 | -1.4142 | -1.4651 | -0.2929 | -0.5412 | -1.4142 |
| **$\lambda_{25}$=270** | 0.5412 | 1.4142 | 0.7654 | -1.0000 | -1.3066 | -1.4142 | -1.8478 | -1.0000 | -0.5412 | -1.4142 |
| **$\lambda_{26}$=281.25** | 0.9239 | 1.4142 | 1.3066 | -0.7071 | -1.6892 | -1.4142 | -1.8478 | -1.7071 | -0.9239 | -1.4142 |
| **$\lambda_{27}$=292.5** | 1.4651 | 1.4142 | 1.6892 | 0.0000 | -1.6892 | -1.4142 | -1.4651 | -2.0000 | -1.4651 | -1.4142 |
| **$\lambda_{28}$=303.75** | 1.8478 | 1.4142 | 1.6892 | 0.7071 | -1.3066 | -1.4142 | -0.9239 | -1.7071 | -1.8478 | -1.4142 |
| **$\lambda_{29}$=315** | 1.8478 | 1.4142 | 1.3066 | 1.0000 | -0.7654 | -1.4142 | -0.5412 | -1.0000 | -1.8478 | -1.4142 |
| **$\lambda_{30}$=326.25** | 1.4651 | 1.4142 | 0.7654 | 0.7071 | -0.3827 | -1.4142 | -0.5412 | -0.2929 | -1.4651 | -1.4142 |
| **$\lambda_{31}$=337.5** | 0.9239 | 1.4142 | 0.3827 | 0.0000 | -0.3827 | -1.4142 | -0.9239 | 0.0000 | -0.9239 | -1.4142 |
| **$\lambda_{32}$=348.75** | 0.5412 | 1.4142 | 0.3827 | -0.7071 | -0.7654 | -1.4142 | -1.4651 | -0.2929 | -0.5412 | -1.4142 |
| | *0.541≤S≤ 1.848* | *S=1.414* | *0.383≤S≤ 1.689* | *-1.0≤S≤ 1.0* | *-1.689≤S≤ -0.383* | *S=-1.414* | *-1.848≤S≤ -0.541* | *-2.0≤S≤ 0.0* | *1.848≤S≤ 0.541* | *S=-1.414* |

*Note: lab tests may be possible for a specified spectrum of polarization states by utilizing the existing Bell set up (Figure 1).*

### *A.2.3 Bell Inequalities - Entire Population*

Table A.6 showed the mean expected values (equation A-6b) over all polarization states (population). Table A.7 showed the predicted Bell inequalities (equation A-6a) for the system (over $360^0$ hemisphere.

**Table A.6**

Mean expected values, $\bar{E}\left(\vec{a}_i, \vec{b}_j\right)$, over all polarization states.

| $a_i/b_j$ | ***$b_1$* (0)** | ***$b_2$* (11.25)** | ***$b_3$* (22.5)** | ***$b_4$* (33.75)** | ***$b_5$* (45)** | ***$b_6$* (56.25)** | ***$b_7$* (67.5)** | ***$b_8$* (78.75)** | ***$b_9$* (90)** | .. | ***$b_{32}$* (348.75)** |
|---|---|---|---|---|---|---|---|---|---|---|---|
| ***$a_1$* (0)** | 1 | 0.913 | 0.686 | 0.356 | -0.029 | -0.41 | -0.728 | -0.935 | -1 | | 0.935 |
| ***$a_2$* (11.25)** | 0.913 | 0.976 | 0.89 | 0.669 | 0.347 | -0.029 | -0.401 | -0.711 | -0.913 | | 0.711 |
| ***$a_3$* (22.5)** | 0.686 | 0.89 | 0.958 | 0.881 | 0.67 | 0.356 | -0.012 | -0.378 | -0.686 | | 0.378 |
| ***$a_4$* (33.75)** | 0.356 | 0.669 | 0.881 | 0.958 | 0.891 | 0.686 | 0.378 | 0.012 | -0.356 | | -0.012 |
| ***$a_5$* (45)** | -0.029 | 0.346 | 0.669 | 0.89 | 0.976 | 0.913 | 0.711 | 0.401 | 0.029 | | -0.401 |
| ***$a_6$* (56.25)** | -0.41 | -0.029 | 0.356 | 0.686 | 0.913 | 1 | 0.935 | 0.728 | 0.41 | | -0.728 |
| ***$a_7$* (67.5)** | -0.728 | -0.401 | -0.012 | 0.378 | 0.711 | 0.935 | 1 | 0.944 | 0.728 | | -0.944 |
| ***$a_8$* (78.75)** | -0.935 | -0.711 | -0.378 | 0.012 | 0.401 | 0.728 | 0.944 | 1 | 0.935 | | -1 |
| ***$a_9$* (90)** | -1 | -0.913 | -0.686 | -0.356 | 0.03 | 0.41 | 0.728 | 0.935 | 1 | | -0.935 |
| .. | | | | | | | | | | | |
| ***$a_{32}$* (348.75)** | 0.935 | 0.711 | 0.378 | -0.012 | -0.401 | -0.728 | -0.944 | -1 | -0.935 | | 1 |

**Table A.7**

Bell Inequalities over the entire $360^0$ hemisphere (polarization population) for several test cases. *(Bell inequality, -2≤S≤2, does not holds)*

| *Test* | *(a,b,a',b')* | *S* | *Test* | *(a,b,a',b')* | *S* |
|---|---|---|---|---|---|
| ***1*** | (a1,b2,a3,b4) = 11.25 | 2.3281 | 6 | (a1,b7,a13,b19) = 67.5 | -2.7941 |
| ***§2*** | **(a1,b3,a5,b7) = 22.5** | 2.7941 | 7 | (a1,b8,a15,b22) = 78.75 | -2.4049 |
| ***3*** | (a1,b4,a7,b10) = 33.75 | 2.0467 | 8 | (a1,b9,a17,b25) = 90 | -2.0000 |
| ***4*** | (a1,b5,a9,b13) = 45 | -0.0587 | 9 | (a1,b10,a19,b28) = 101.25 | -2.3281 |
| ***5*** | (a1,b6,a11,b16) = 56.25 | -2.0786 | 10 | (a1,b11,a21,b31) = 112.5 | -2.7941 |

*§This prediction showed a complete agreement with several lab tests in Tables 2 and 3.*

# Appendix B: Perspectives on Intrinsic Properties

Perspectives on intrinsic properties are from well established photon diffraction experiments *(particles)* and a biological system *(athletes)* analogous to Bell test on correlated photons in visible spectrum. The perspectives are align with this study "on quantum entanglement and nonlocality."

## *Particles*

Let us review this very basic optical diffraction result first. Suppose light passes through a double slit of 0.01 *mm* or through diffraction gratings with 1000 *lines per centimeter.* Photons are being detected on the screen at 2.0 *m* from the slit. For this illustration, only 3 wavelengths in the visible spectrum are shown. The prediction steps and detection results are given in Table B.1. Figure B.1 shows visible light spectrum from diffraction experiment and interference pattern (emerging properties) from slits experiment. As a photon passes, it spreads as a function of wavelength *(λ)* and slit size *(d)*.

In a slits experiment, things become a little interesting, suppose we are sending one photon at a time and recording the detection. The results will look random initially due to the limited number of data points, after which interference patterns will appear. Such interference patterns can easily be described by quantum mechanics wavefunction superposition. Let us conclude this simple test with the following remark: what happens if we apply the computation steps in Table B.1 by incorporating the initial details of each photon (such as quantized polarization, momentum states vectors when exiting the slits and no further disturbances afterward)? Can we obtain the same prediction as given by wavefunction superposition? The case study given next will shed some light on this question.

**Table B.1**
Basic light diffraction experiment.

| Experiment Setup | $J_i$ | $\lambda_i$ *(nm)* | $\theta_i = \sin^{-1}(J_i \lambda_i / d)$ *(degree)* | $y_i = x \tan(\theta_i)$ *(cm)* |
|---|---|---|---|---|
| 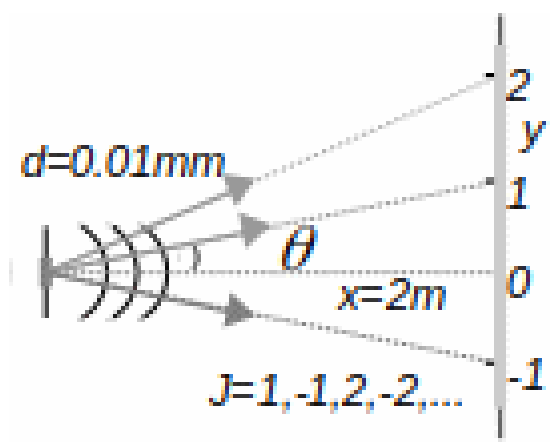  | 1 | 485 *(blue)*<br>565 *(green)*<br>750 *(red)* | 2.78<br>3.24<br>4.30 | 9.71<br>11.32<br>15.04 |
| | 2 | 485<br>565<br>750 | 5.57<br>6.49<br>8.63 | 19.49<br>22.75<br>30.34 |

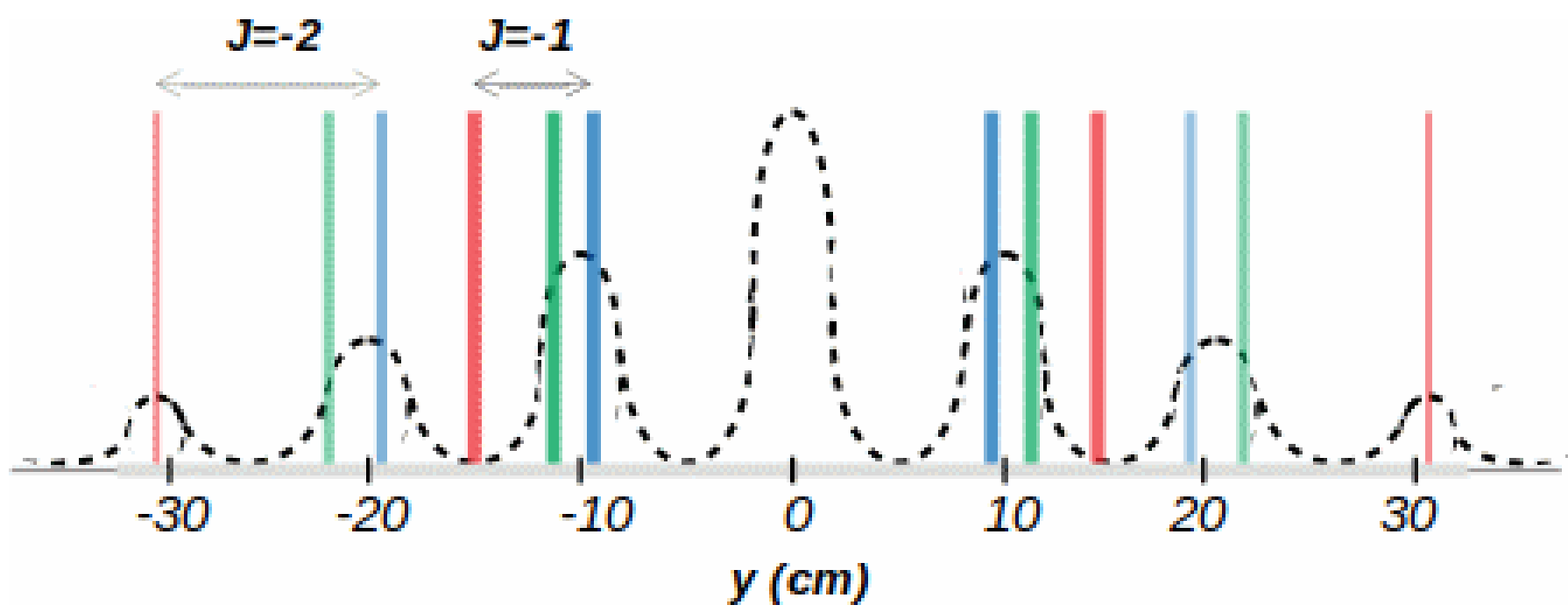

**Figure B.1.** Schematic showing the visible light spectrum from diffraction experiment and an interference pattern (emerging properties) from slits experiment. This is a perspective view on intrinsic nature (*such as wave length, spin, momentum*) and emerging properties of a given observation system.

*Athletes*

The goal in this example is to illustrate how an intrinsic property for a physical system appears to an observer as nonlocal. The physical interpretation given by wavefunction superposition for many optical entanglements may not be a complete description.

Let us first define a physical system. A system can be any elementary particle, molecule or biological species where each system has an individual identity. When a person such as an athlete is trained for a given race, the person is conditioned on his or her training program, which ultimately determines the expected outcomes. The variance in the expected outcomes is much narrower for a highly trained professional than for nonprofessional athletes.

With great curiosity, this study analyzed the performance completion times of 26,260 trained athletes from a highly recognized annual event (Boston Marathon 2023). Among these participants, there were 14,911 males and 11,349 females. All the participants were highly trained athletes who were challenging themselves (i.e., aiming for personal best). The statistical analysis of the participants' performance on age and sex revealed some interesting correlations that were very similar to the light diffraction interference patterns.

Table B.2 summarizes the statistical analysis of the finishing times for several age groups. Let us check the first age group (18-39); here, a total of 5,285 men had an average finish time of $\mu$=196.50 *minutes* with a standard deviation of $\sigma$=42.50; 4,884 women had an average finish time of $\mu$ =227.91 *minutes* with a standard deviation of $\sigma$=42.76. The distribution was slightly left skewed if we compared the median with the mean value. Figure B.2 shows the mean values for all age groups only as a line diagram. This result looks very similar to the interference patterns in double-slit experiments. Interestingly, the overall time difference between men and women was 25 *min*, which was close to the 30 *min* set in the qualifying entry. The first age group was 4.2 times wider than the other age groups. All the lead athletes were within this age group (18-39). Certainly, there should be a more visible spectrum (similar to Figure B.2) if we further analyze the distribution of this group. A photon experiment on "a delayed choice quantum eraser" by Kim *et al.* (2000) may be analogous to this case study. This is a good illustration of Bell's characterization of photon polarization states (state vector at individual instance vs all states vectors over the entire $360^0$ hemisphere).

In conclusion, a physical reality (such as intrinsic properties) may appear visible by repeating a large number of tests if such properties have any weighted interaction with the observation setup. One may think what this case study has anything to do with the hidden variables (i.e., intrinsic properties) we are dealing with in physics. Einstein believes that every physical reality of any given system can be predicted with 100% certainty by a complete theory. An attempt here is to demonstrate that the nonlocality arguments on entanglements observed in many experiments are inconclusive.

**Table B.2**
Statistical analysis of the completion time of the highly conditioned athletes (data from a recognized annual event).

| ***Age Group*** | ***Men*** | | | | ***Women*** | | | |
|---|---|---|---|---|---|---|---|---|
| | *N* | *μ (min)* | *σ* | *Median* | *N* | *μ (min)* | *σ* | *Median* |
| ***18-39*** | 5285 | 196.50 | 42.50 | 179.60 | 4884 | 227.91 | 42.76 | 212.38 |
| ***40-44*** | 2241 | 203.00 | 39.15 | 189.28 | 1878 | 230.76 | 37.57 | 219.21 |
| ***45-49*** | 2252 | 210.26 | 37.09 | 198.97 | 1817 | 235.87 | 33.18 | 227.17 |
| ***50-54*** | 2052 | 221.32 | 40.38 | 207.39 | 1207 | 244.61 | 35.03 | 233.92 |
| ***55-59*** | 1422 | 228.29 | 37.27 | 217.36 | 815 | 253.28 | 35.01 | 244.13 |
| ***60-64*** | 1110 | 239.82 | 36.79 | 229.95 | 527 | 258.9 | 31.50 | 252.77 |
| ***65-69*** | 549 | 252.95 | 37.06 | 242.58 | 221 | 272.10 | 30.34 | 267.75 |

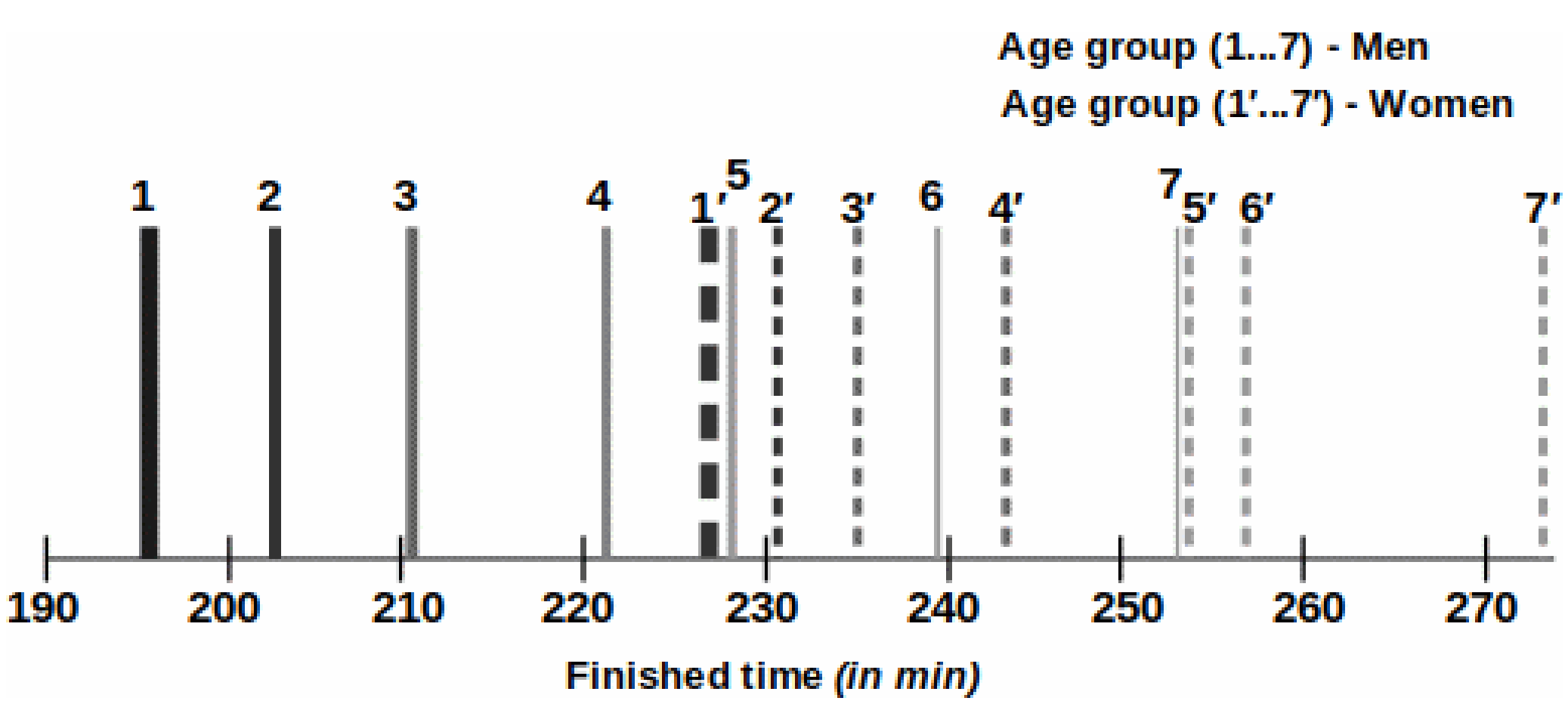

**Figure B.2.** Schematic showing the mean completion times by age group and sex (highlighting performance status). This is a perspective view on emerging outcomes of a large number of participants in a given race (statistically unbiased).

Test examples from single-photon diffraction experiments and the performances of a large number of athletes at an event clearly showed how the intrinsic properties appeared as a visible spectrum in the observations. An overall population response here depends on the relative positions of the particles (particles system) or athletes (sports event) by their intrinsic properties. The Bell inequalities on local causality are restricted to each individual instances (such as photon polarization) but not to the entire population. This is a good illustration that the local causality is likely correct and the quantum mechanics prediction on a unified system is an approximation (not physical).